\documentstyle[12pt,epsf,epsfig,psfig]{article}
\def\J{$J/\psi$}

\def\x{\chi}
\def\P{$\psi'$}

\def\U{$\Upsilon$}

\def\C{c{\bar c}}

\def\B{b{\bar b}}

\def\e{\epsilon}

\def\S{\sigma_{\C}}

\def\be{\begin{equation}}
\def\ee{\end{equation}}

\def\lsim{\raise0.3ex\hbox{$<$\kern-0.75em\raise-1.1ex\hbox{$\sim$}}}
\def\gsim{\raise0.3ex\hbox{$>$\kern-0.75em\raise-1.1ex\hbox{$\sim$}}}

\def\NP{{ Nucl.\ Phys.\ }}
\def\PL{{ Phys.\ Lett.\ }}
\def\PR{{ Phys.\ Rev.\ }}

\def\PRL{{ Phys.\ Rev.\ Lett.\ }}

\def\ZP{{ Z.\ Phys.\ }}

\begin{document}

\noindent July 19, 2000~ \hfill BI-TP 2000/26

\vskip 1.5 cm

\centerline{\Large{\bf Phase Transitions in QCD}\footnote{Invited talk
at the 3rd Catania Relativistic Ion Studies {\it CRIS 2000}, 
Acicastello/Italy, May 22 - 26, 2000; at the XL Cracow School of 
Theoretical Physics, Zakopane/Poland, June 3 -11, 2000; and at the
Conference on Strong and Electroweak Matter {\it SEWM 2000}, Marseille/France,
June 13 - 17, 2000.}}

\vskip 1.0cm

\centerline{\bf Helmut Satz}

\bigskip

\centerline{Fakult\"at f\"ur Physik, Universit\"at Bielefeld}
\par
\centerline{D-33501 Bielefeld, Germany}

\vskip 1.0cm

\noindent

\centerline{\bf Abstract}

\medskip

At high temperatures or densities, hadronic matter shows different
forms of critical behaviour: colour deconfinement, chiral symmetry
restoration, and diquark condensation. I first discuss the conceptual
basis of these phenomena and then consider the description of colour
deconfinement in terms of symmetry breaking, through colour screening
and as percolation transition.

\vskip 1.5cm

\noindent{\large 1.\ States of Matter}

\bigskip

Hadronic matter is endowed with an inherent density limit. The usual
hadrons have an intrinsic size $r_h \simeq 1$ fm, so that a hadron
needs a space of volume $V_h \simeq (4\pi/3)~r_h^3$ to exist. Therefore
\be
n_c \simeq (1/V_h) \simeq 0.25~{\rm fm}^{-3} \simeq 1.5~n_0 \label{1.1}
\ee
is the limiting density for such a medium; here $n_0 = 0.17$ fm$^{-3}$
denotes the density of normal nuclear matter. In turn, this also leads
to a limiting temperature for hadronic matter,
\be
T_c \sim (1/r_h) \sim 0.2~{\rm GeV}, \label{1.2}
\ee
as pointed out by Pomeranchuk almost fifty years ago \cite{Pom51}.
Considering an ideal gas of resonances, whose composition was based on a
classical partitioning problem \cite{Euler}, Hagedorn found a very
similar limit, which he proposed as the ultimate temperature of strongly
interacting matter \cite{Hagedorn}. Dual resonance dynamics led to an
essentially equivalent composition law \cite{Dual}. Soon afterwards
Cabibbo and Parisi noted that the `limit' more likely corresponded to a
critical point, signalling the transition to a new state of matter, the
quark-gluon plasma \cite{C-B}. What remains is the realization that, on
geometric, combinatoric or dynamic grounds, hadron thermodynamics
defines its own limit.

\medskip

This limit can be approached in two ways. The obvious is to compress
cold nuclear matter beyond $n_0$; but in relativistic thermodynamics,
`heating' mesonic matter leads to particle production and thus also
increases the density. As a result, strongly interacting matter has a
$T\!-\!\mu_B$ phase diagram, where $T$ denotes the temperature and
$\mu_B$ the chemical potential specifying the overall baryon number
density. In the $T\!-\!\mu_B$ plane, there must thus be a limiting
curve for hadronic matter, beyond which the density is too great to
allow the existence of hadrons.

\medskip

Since QCD defines hadrons as bound states of quarks, the general phase
structure of strongly interacting matter is quite evident: for densities
below $n_c$, it consists of colourless hadrons, i.e., colour singlet
bound states of three quarks or of a quark-antiquark pair. Above $n_c$,
deconfinement leads to a medium consisting of coloured constituents.
There are three possible forms for such constituents:
\begin{itemize}
\vspace*{-0.3cm}
\item{coloured massive quark-gluon states: constituent quarks;}
\vspace*{-0.3cm}
\item{coloured massive quark-quark states: diquarks;}
\vspace*{-0.3cm}
\item{coloured massless quarks and gluons: a quark-gluon plasma.}
\vspace*{-0.3cm}
\end{itemize}
\noindent
What actually happens in the different regions of the $T\!-\!\mu_B$
diagram?

\medskip

Before addressing this question, we have to consider how deconfinement
can take place. The confining potential between a static quark and
antiquark separated by a distance $r$ has the idealized form
\be
V(r) \sim \sigma r, \label{1.3}
\ee
where $\sigma \simeq 0.8$ GeV/fm is the string tension. The quarks
inside a hadron are therefore confined: the hadron cannot be broken up
into its coloured constituents, since this would require an infinite
amount of energy.

\medskip

In a dense medium, however, there is another way to dissociate bound
states. The presence of many other charges leads to charge screening,
which reduces the range of the forces between charges. A well-known
example is Debye screening, which suppresses the long-range part of the
Coulomb potential between electric charges,
\be
V(r) = {1 \over r} \to {1 \over r}~e^{-\mu r},
\label{1.4}
\ee
where $r_D=\mu^{-1}$ is the Debye radius, defining the range of the
force remaining effective between charges in the medium. When it becomes
smaller than the atomic binding radius, an insulator consisting of
charge-neutral atoms turns into a conducting plasma of unbound electric
charges \cite{Mott}. In QCD, the corresponding effect leads to colour
screening,
\be
V(r) = \sigma r  \to \sigma r \left[ {1 - e^{-\mu r} \over \mu r}
\right], \label{1.5}
\ee
where $\mu^{-1}$ now defines the colour screening radius\footnote{The
difference in the form of the screening functions in Eq.\ (\ref{1.4})
and (\ref{1.5}) is due to the different forms of the unscreened
potentials \cite{Dixit}.}. Deconfinement thus is the insulator-conductor
transition of QCD, with colourless bound states as constituents below
and coloured constituents above the deconfinement point. But what is
the nature of the conducting phase here?

\medskip

One way to study that is to consider the effective quark mass. The
input quark masses in the QCD Lagrangian are (for $u$ and $d$ quarks)
almost zero, $m_q \simeq 0$. In the confined phase, hadrons behave as if
they consist of constituent quarks of mass $m_Q$, with $m_n \simeq 3
m_Q$ and $m_{\rho} \simeq 2 m_Q$, for nucleons and (non-Goldstone)
mesons, respectively. Hence here the quarks manage to `dress'
themselves with gluons to acquire a mass $m_Q\simeq0.3$ GeV.

\medskip

At sufficiently high temperatures, thermal motion will presumably
`shake off' the dressing, so that somewhere in the course of the
hadron-quark matter transition there will be an effective quark mass
shift $m_Q \to m_q$. For vanishing $m_q$, the QCD Lagrangian is
chirally symmetric; hence this chiral symmetry must be spontaneously
broken in the confined phase and restored in the hot QGP. One therefore
often refers to the shift in effective quark mass $m_Q \to 0$ as chiral
symmetry restoration. Such a shift is more general, however, and can
occur as well for $m_q\not= 0$, as shown by the shift of the effective
electron mass between insulator and conductor.

\medskip

That leads to the next question: is there also a colour superconductor?
At low temperatures, collective effects of an electrically conducting
medium can overcome the Coulomb repulsion between like charges and lead
to a binding of electrons into doubly charged Cooper pairs. These,
being bosons, can condensate to form a superconductor. In QCD, the
conditions for creating a superconductor are in fact much more
favorable. An attractive local potential couples two triplet quark
states to a bosonic anti-triplet diquark state, so that in QCD there is
a dynamical basis for colour superconductivity through diquark
condensation \cite{Bailin}.

\medskip

In very recent years, the low temperature, high baryon density part of
the QCD phase diagram has received much renewed attention, resulting in
the prediction of  different superconducting phases and several
transitions \cite{Krishna}. Although of great theoretical interest,
this region is for the time being accessible neither to lattice studies
nor to experiment. I shall therefore concentrate here on the high
temperature, low baryon density region and refer to \cite{Krishna} for
a discussion of colour superconductivity.

\medskip

\begin{figure}[htp]
\centerline{\psfig{file=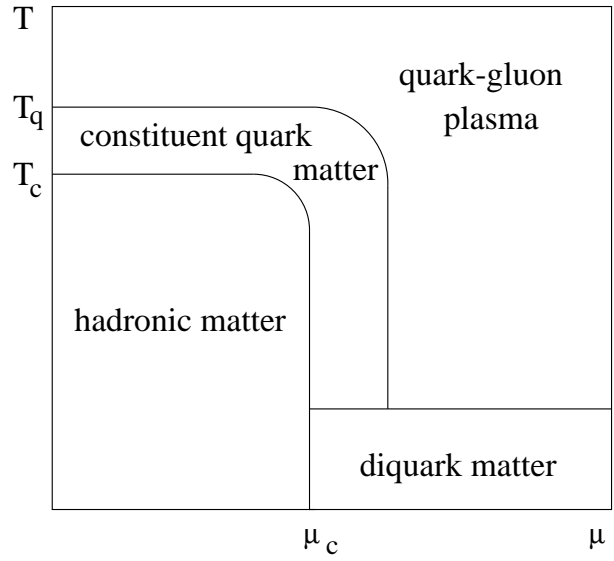,width=8cm}}
\vskip0.3cm
\caption{Four-phase diagram of strongly interacting matter.}
\label{F1.1}
\end{figure}

Taking into account what was said, a first guess of the QCD phase
diagram leads to a four-phase structure of the generic form shown Fig.\
\ref{F1.1}; as noted, the `diquark phase' shown there may well consist
of several different superconducting phases. In any case, however,
lattice QCD tells us that Fig.\ \ref{F1.1} is wrong: at $\mu=0$,
deconfinement and chiral symmetry restoration coincide, so that there
is no constituent quark phase. One of the main points I want to
address here is why this is so.

\medskip

\medskip

\begin{figure}[htp]
\centerline{\psfig{file=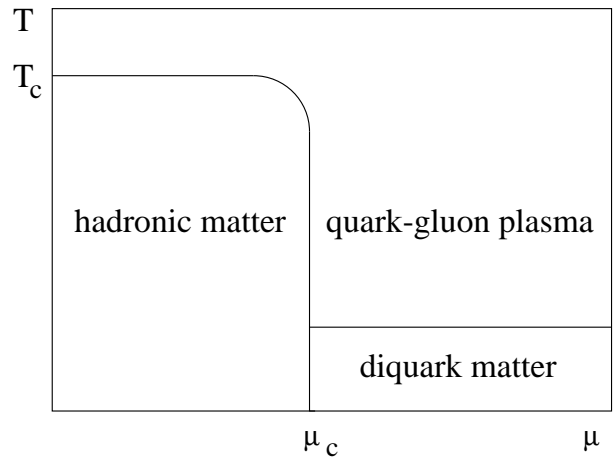,width=8cm}}
\vskip0.3cm
\caption{Three-phase diagram of strongly interacting matter.}
\label{F1.2}
\end{figure}

A second guess is shown in Fig.\ \ref{F1.2}, with hadronic, diquark and
QGP phases. As far as we know, this one may well be correct; for
$\mu_B=0$, it is, as we shall see from the results provided by finite
temperature lattice QCD.

\medskip

\medskip

In the classical study of critical behaviour, the analyticity of
the partition function $Z(T,\mu,V)$ in the thermodynamic limit of
infinite volume, $V \to \infty$, governs the phase structure, with
critical points defined through the divergence of derivatives of
$Z(T,\mu,V)$. The occurrence of such singularities can be attributed to
different underlying physics mechanisms. In statistical physics,
spontaneous symmetry breaking, charge screening and cluster
percolation have received particular attention. In the following
sections, I shall consider these three `mechanisms' in the context of
statistical QCD.

\bigskip

\noindent{\bf \large 2.\ Symmetry Breaking}

\bigskip

In contrast to the statistical mechanics of condensed atomic matter, the
phase structure of strongly interacting matter can be determined {\sl ab
initio} from QCD as input dynamics, without intermediate models. Given
the QCD Lagrange density
\be
{\cal L}_{\rm QCD} = -F_{\mu \nu} F^{\mu \nu} - {\bar \psi}
(i{\gamma^{\mu} \partial_{\mu}} - g {\gamma^{\mu} A_{\mu}} + m_q) \psi,
\label{2.1}
\ee
where the first term describes the pure gluon sector and the second the
quark-gluon interactions, one defines the partition function
at temperature $T$ as
\be
Z(T,V) = \int dA~d{\bar \psi}~d{\psi}~\exp\{- S({\cal L},T) \}.
\label{2.2}
\ee
Here $A$ denotes the gluon and $\psi$ the quark fields; the QCD action
is given by \cite{Bernard}
\be
S({\cal L};T,V) = \int_V d^3x \int_0^{1/T} d\tau~ {\cal
L}(A(x,\tau),\psi(x,\tau)),
\label{2.3}
\ee
as integral over the volume $V$ of the system and a slice of
thickness $1/T$ in the imaginary time $\tau=ix_0$. From $Z(T,V)$ one
then obtains the usual thermodynamic functions; the derivative with
respect to $T$ leads to the energy density $\e(T)$, that with respect to
$V$ the pressure $P(T)$, and so on. Both the dynamic input theory and
its associated thermodynamics are thus completely specified; the problem
lies in the evaluation, for which one has to resort to the numerical
simulation \cite{Creutz} of the lattice formulation \cite{Wilson}.

\medskip

The conventional deconfinement probe in finite temperature lattice QCD
is the expectation value $L(T)$ of the Polyakov loop \cite{MS,KPS}.
Through
\be
L(T) \sim \lim_{r \to \infty} e^{-V(r)/T},
\label{2.4}
\ee
it is related to the potential $V(r)$ coupling a static quark-antiquark
pair. In the confined phase, this potential diverges as $V(r)$ for
$r \to \infty$ (see Eq.\ (\ref{1.3})), while in the deconfined phase it
converges to a finite value. It is found that $L(T)=0$ in the
temperature range $T \leq T_c$, defining the confinement region, and
that $L(T) > 0$ for $T > T_c$, specifying the deconfinement region.

\medskip

Strictly speaking, $V(r) \sim \sigma r$ diverges for $r \to \infty$
only in a theory with infinitely heavy quarks. In real QCD with light
dynamical quarks, the string breaks when it becomes energetically more
favourable to produce a quark-antiquark pair, i.e., when $V(r) \simeq 2
m_Q$, where $m_Q$ is the mass of a `dressed' constituent quark and
$2m_Q$ the mass of a (non-Goldstone) meson. The quark of this newly
produced pair combines with the original antiquark, the antiquark with
the original quark, thus making two strings out of one. In full QCD,
the Polykov loop therefore does not vanish in the confining region, but
only becomes exponentially small,
\be
L(T) \simeq e^{-2m_Q/T} ~~~{\rm for}~~T\leq T_c.
\label{2.5}
\ee
For $m_Q \simeq 0.3$ GeV and $T_c \simeq 0.15$ GeV, this makes $L(T_c)
\simeq 0.02$ instead of zero. Nevertheless, $L(T)$ is now no longer a
real order parameter, and deconfinement therefore seems to be not really
defined as a critical phenomenon in QCD with light dynamical quarks.
We shall return to this problem several times and show how it might be
solved.

\medskip

From statistical mechanics, it is known that phase transitions are
generally associated to symmetries of the system. Thus the Hamiltonian
of the simplest spin theory, the Ising model, is
\be
{\cal H} = -J \sum_{i<j} s_i s_j - B \sum_i s_i,~~~s_i=\pm 1~\forall~
i,
\label{2.6}
\ee
where the first sum runs only over nearest neighours on the lattice,
$J$ denotes the exchange energy between spins and $B$ an external
magnetic field. For $B=0$, $\cal H$ is invariant under the global $Z_2$
symmetry of flipping all spins, $s_i \to -s_i~\forall~ i$. The
thermodynamic states of this system share this symmetry for $T \geq
T_c$, where $T_c$ now is the Curie point; as a result, the expectation
value of the spin, the magnetization $m(T)$, vanishes in this
`disordered' temperature region. Below $T_c$, however, the system
becomes ordered, the spins choose to align either up or down, making
$m(T) \not= 0$. Since `up' or `down' are equally likely, the symmetry as
such is preserved; the actual state of the system, however,
spontaneously breaks it by choosing one or the other. The magnetization
transition of the Ising model thus corresponds to the spontaneous
breaking of the inherent global $Z_2$ symmetry of the Ising Hamiltonian.

\medskip

This line of argument applies directly also to deconfinement in pure
$SU(N)$ gauge theories \cite{MS,KPS}. The corresponding Lagrangian
in lattice QCD
\be
{\cal L}_{\rm SU(N)}(U_{ik}U_{kl}U^+_{lm}U^+_{mi})
\label{2.7}
\ee
depends on the products of four $SU(N)$ matrices $U_{ij}$ on the links
of the smallest closed loops of the lattice. It remains invariant under
a global `flip' $z_N \in Z_N \subset SU(N)$ of all matrices associated
to a spatial hyperplane, with $U_{x,\tau} \to z_N U_{x,\tau}~\forall~x$;
here $z_N=\exp \{r(2\pi i/N)\}$ with $r=1,2,...,N$.

\medskip

The Polyakov loop, on the other hand, does not remain invariant under
such global $Z_N$ transformations, with
\be
L ~ \sim~ < Re~Tr~\Pi_{\tau=1}^{N_{\tau}} U_{x,\tau}> ~\to~ z_N~L.
\label{2.8}
\ee
It is thus the analogue of the magnetization of the Ising model, in the
sense that it tests if the state of the system shares or spontaneously
breaks the symmetry of the Lagrangian. This feature becomes particularly
transparent for $SU(2)$ gauge theory, where $z_2 = \pm 1$, so that the
transformation thus just means flipping the sign of the Polyakov loop,
$L \to - L$. In the temperature region in which $L(T)=0$, i.e., in the
confinement region, the states are $Z_N$-symmetric, while for
deconfinement, with $L(T) > 0$, the $Z_N$ symmetry is spontaneously
broken. In other words, deconfinement in pure $SU(N)$ gauge theory can
be defined as the spontaneous breaking of a global $Z_N$ symmetry of the
corresponding Lagrangian.

\medskip

\begin{figure}[htb]
\mbox{
\hskip1cm\epsfig{file=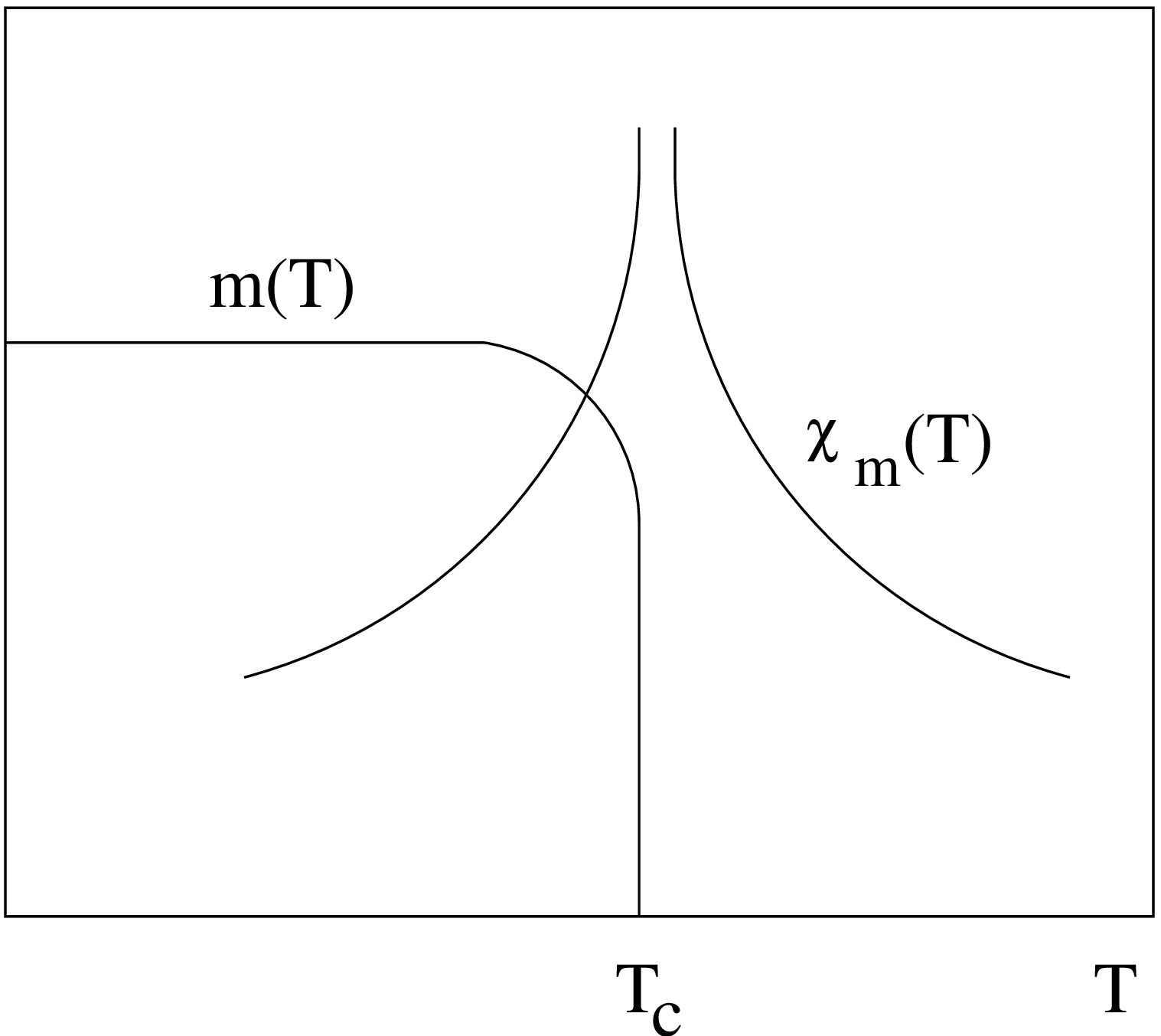,width=6cm}
\hskip1.5cm
\epsfig{file=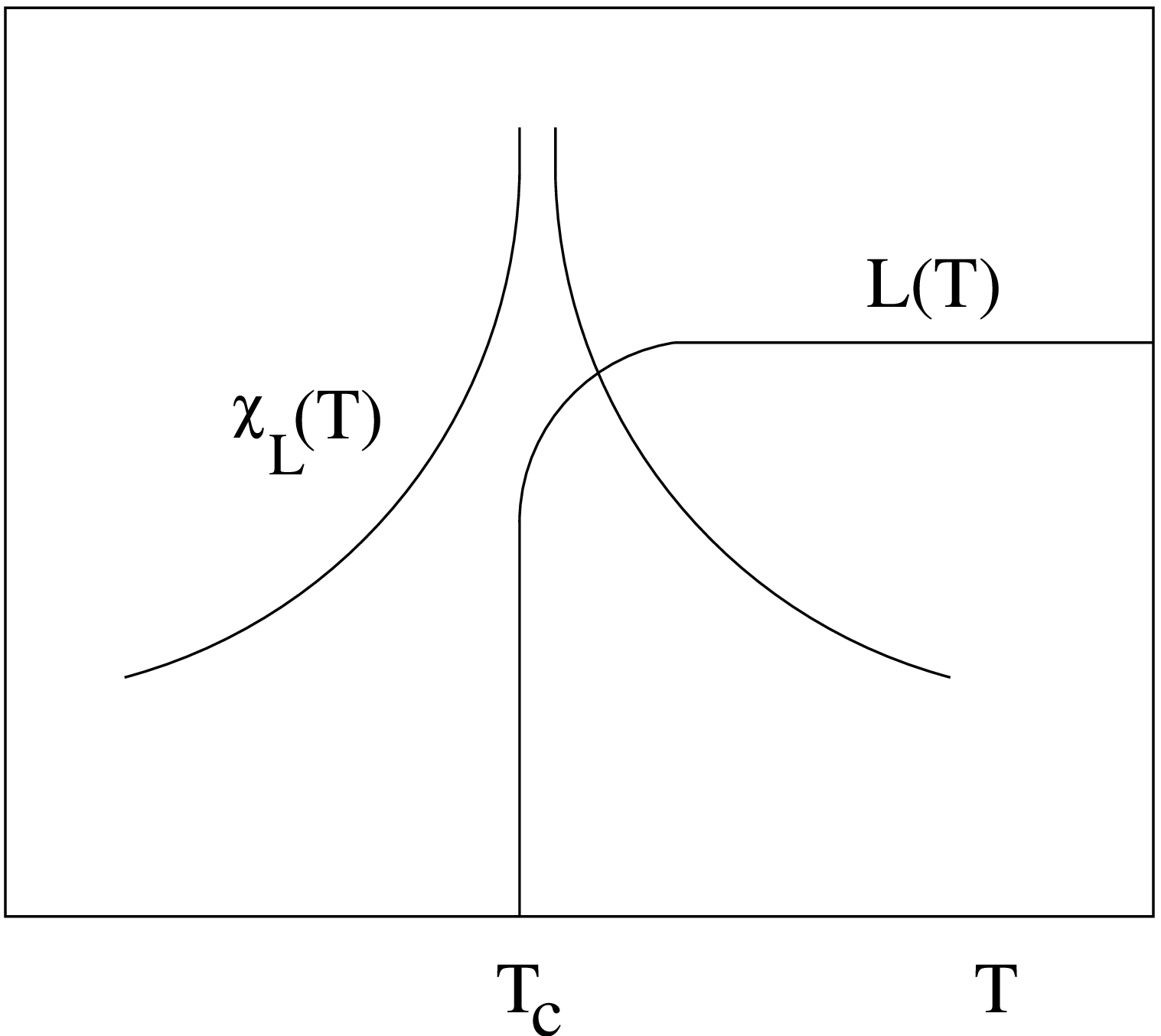,width=6cm}}
\vskip0.3cm
\caption{Schematic temperature dependence of the magnetization $m(T)$ and the
Polyakov loop $L(T)$, together with the corresponding susceptibilities.}
\label{F2.1}
\end{figure}

The similarity between spin and gauge systems goes in fact much further
\cite{SY}. In Fig.\ \ref{F2.1}, we compare schematically the
temperature behaviour of the Polyakov loop $L(T)$ and the magnetization
$m(T)$, together with that of the corresponding susceptibilities
$\x_L (T)$ and $\x_m(T)$. The latter measure the fluctuations of
the order parameters at the transition point and thus diverge there.
For $SU(2)$ gauge theory as well as for the Ising model, the transition is
continuous, and so near $T_c$ the functional behaviour in the two cases
can be written as
\be
L(T) \sim (T-T_c)^{\beta_L},~T>T_c; ~~ \x_L(T) \sim |T-T_c|^{-\gamma_L},
\label{2.9}
\ee
and
\be
m(T) \sim (T_c-T)^{\beta_m},~T<T_c; ~~ \x_m(T) \sim |T-T_c|^{-\gamma_m},
\label{2.10}
\ee
where $\beta$ and $\gamma$ denote the critical exponents for the two
transitions. While $SU(N)$ gauge theories in general have a more complex
structure than spin theories, their critical behaviour becomes in fact
identical: they belong to the same universality class \cite{SY}, which
means that $\beta_L=\beta_m$ and $\gamma_L=\gamma_m$. The
confinement/deconfinement transition in $SU(N)$ gauge theories is thus
structurally the same as the disorder/order transition in spin theories;
both are based on the spontaneous breaking of a global $Z_N$ symmetry of
the underlying dynamics.

\medskip

\begin{figure}[htb]
\centerline{\psfig{file=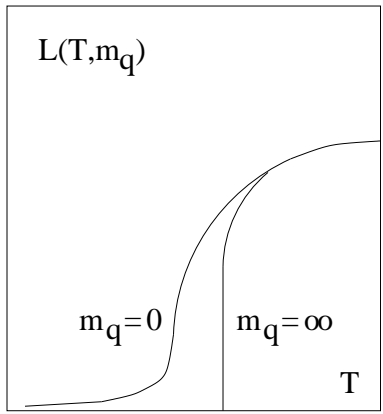,width=5cm}}
\vskip0.3cm
\caption{Schematic temperature dependence of the Polyakov loop in pure $SU(3)$
gauge theory ($m_q=\infty$) and in full QCD with two massless quark
flavours ($m_q=0$).}
\label{F2.2}
\end{figure}

The introduction of dynamical quarks in full QCD explicitly breaks this
$Z_N$ symmetry; it effectively adds a term to the $SU(N)$ action
which is proportional to $L$:
\be
S_{\rm QCD} \sim S_{\rm SU(N)} + \kappa(m_q) L,
\label{2.11}
\ee
where $\kappa(m_q) \to 0$ for $m_q \to \infty$. Comparing Eqs.
(\ref{2.11}) and (\ref{2.6}), we see that dynamical quarks in a sense
play the role of an external field $B$ in spin theory. Just as $B$ aligns
the spins and prevents $m(T)$ from ever completely vanishing, so
does $m_q$ result in a Polyakov loop which is always non-zero.
We had seen above that another way of arriving at this conclusion is
through string breaking. Hence there must be some implicit relation
between the effective external field and the constituent quark mass
determining the string breaking point. The effect of dynamical quarks on
the Polyakov loop is shown schematically in Fig.\ \ref{F2.2}. We note
in particular that even for $m_q=0$, there is `almost critical' behaviour,
with a sharp variation at a temperature considerably below the
deconfinement temperature of pure $SU(3)$ gauge theory What are the
reasons for this behaviour?

\medskip

From the string breaking picture, we expect the effective external
field acting on the Polyakov loop as generalized spin to be inversely
proportional to the constituent quark mass, $B \sim 1/m_Q$
\cite{Gavai,HS-deco,Digal}. Thus deconfinement should occur when $m_Q
\to 0$.

\medskip

\begin{figure}[htb]
\centerline{\psfig{file=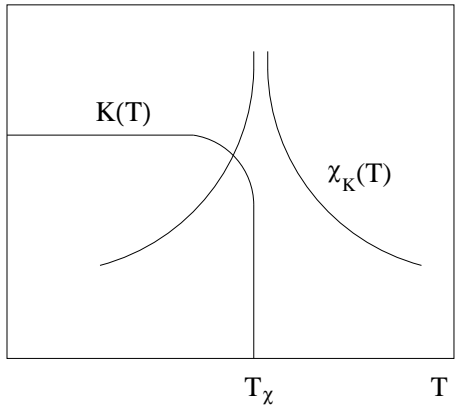,width=6cm}}
\vskip0.3cm
\caption{Schematic temperature dependence of the chiral condensate $K(T)$
in full QCD, together with the corresponding susceptibility.}
\label{F2.2a}
\end{figure}

For $m_q=0$, the QCD Lagrangian is chirally symmetric; however, the
state of the system under given conditions need not share this symmetry.
The chiral condensate, $K(T) \equiv \langle \psi \bar{\psi} \rangle
\sim m_Q^3$, provides an order parameter to probe if and when the chiral
symmetry of the Lagrangian is spontaneously broken. It is found that
\be
K(T) \not= 0 ~~~{\rm implying}~~~m_Q \not=0~~~\forall~T < T_{\x},
\label{2.12}
\ee
and
\be
K(T) = 0 ~~~{\rm implying}~~~m_Q =0~~~\forall~T > T_{\x}.
\label{2.13}
\ee
where $T_{\x}$ MeV is the chiral symmetry restoration temperature.
The functional behaviour of $K$ and the corresponding
fluctuation susceptibility $\x_K$,
\be
K(T) \sim (T_{\x}-T)^{\beta_K},~ T<T_{\x};~~ \x_K(T)
\sim |T-T_{\x}|^{-\gamma_K},
\label{2.13a}
\ee
is illustrated in Fig.\ \ref{F2.2a}.
For $T<T_{\x}$, $K$ is large and hence the effective external field
$B \sim 1/m_Q \sim 1/K(T)$ is small, so that the Polyakov loop is almost
disordered, implying confinement-like behaviour. At $T=T_{\x}$, $B$
suddenly becomes large; it aligns the Polyakov loops, implying the onset
of deconfinement. We thus find that chiral symmetry restoration induces
colour deconfinement.

\medskip

\begin{figure}[htp]
\begin{center}
\epsfig{file=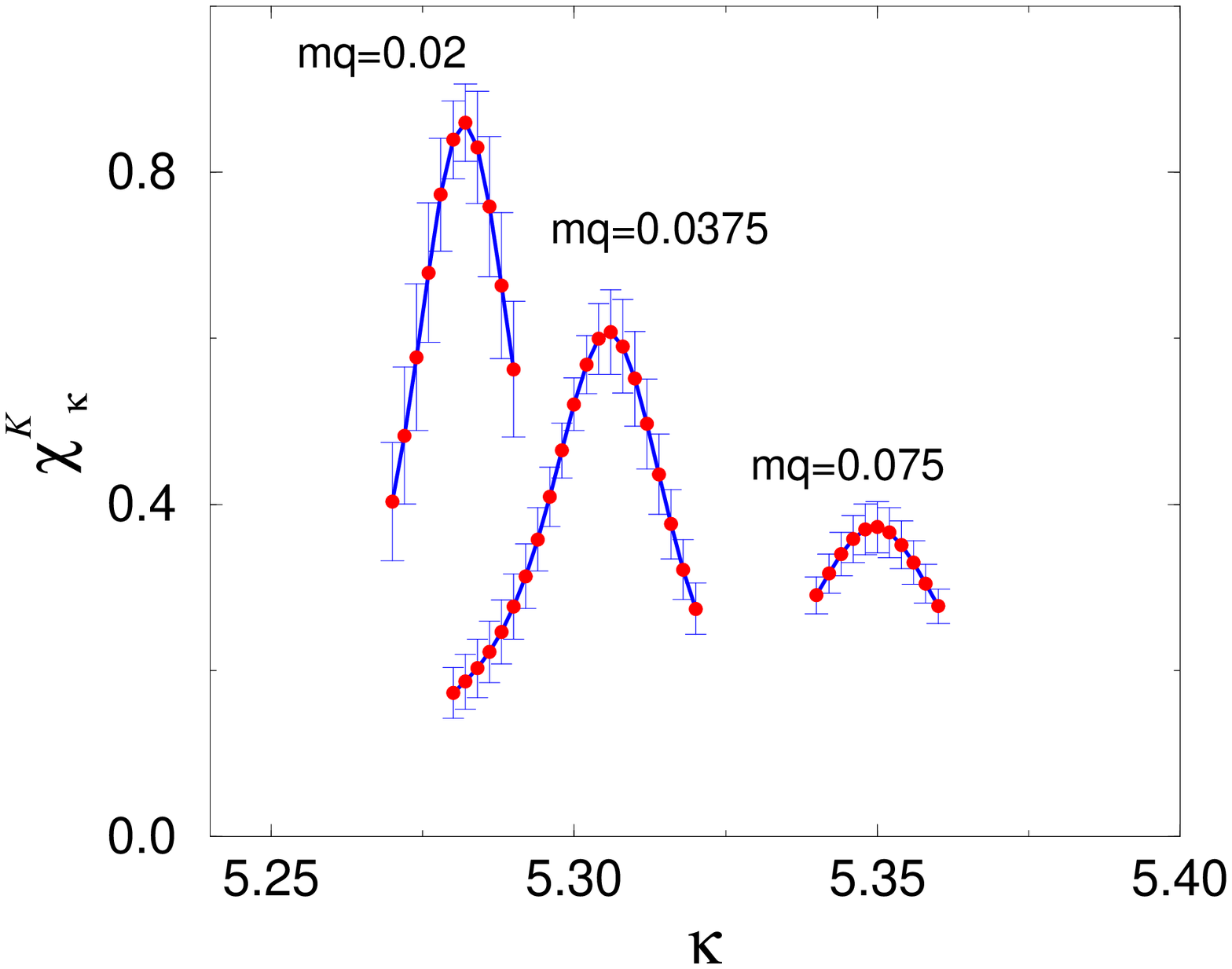,width=7cm}
\hskip0.5cm
\epsfig{file=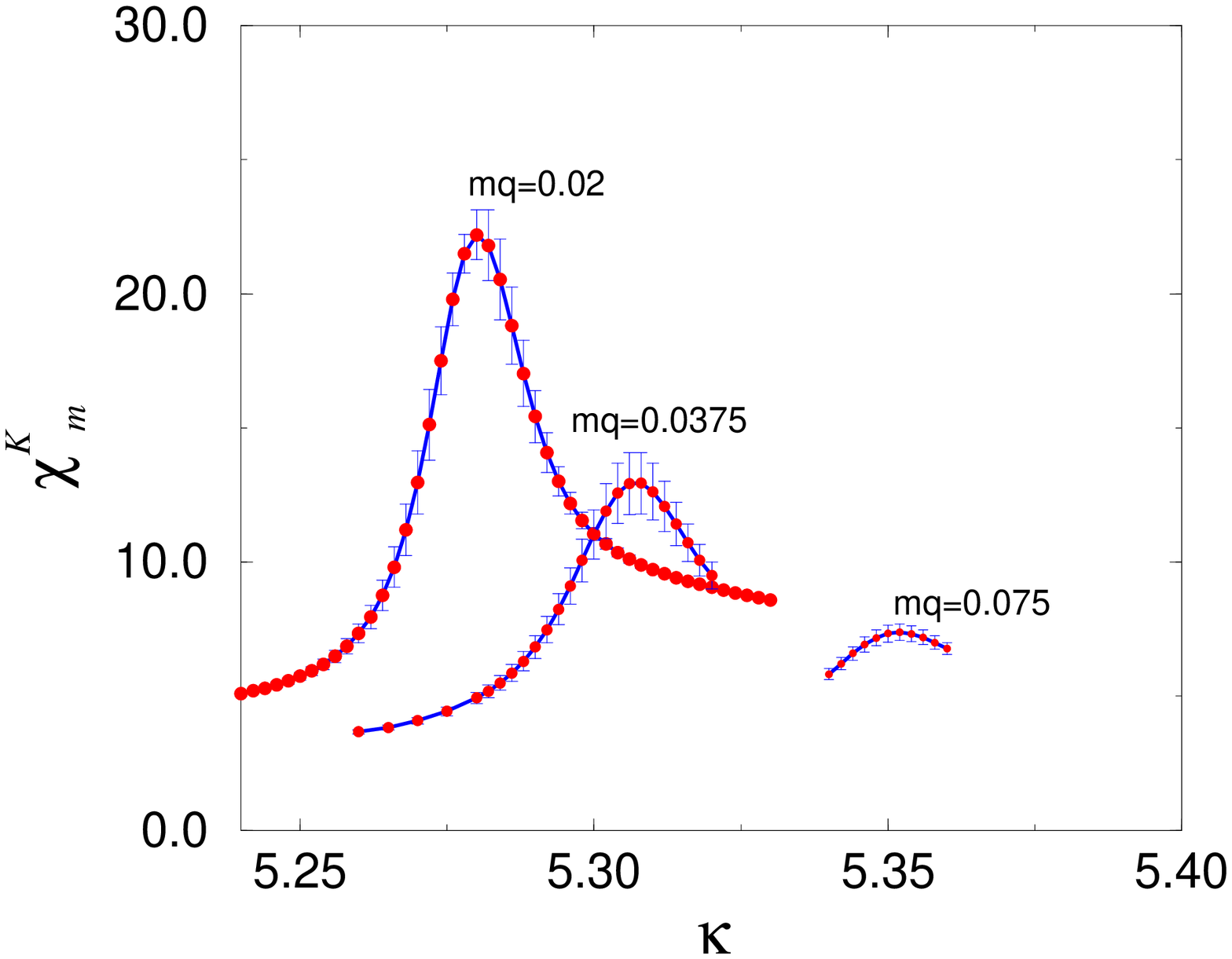,width=7cm}
\end{center}

\caption{The chiral susceptibilities $\x^K_{\kappa}$ and $\x^K_{\kappa}$
as functions of the temperature variable $\kappa=6/g^2$.}
\label{F2.3a}
\end{figure}

\begin{figure}[htb]
\begin{center}
\epsfig{file=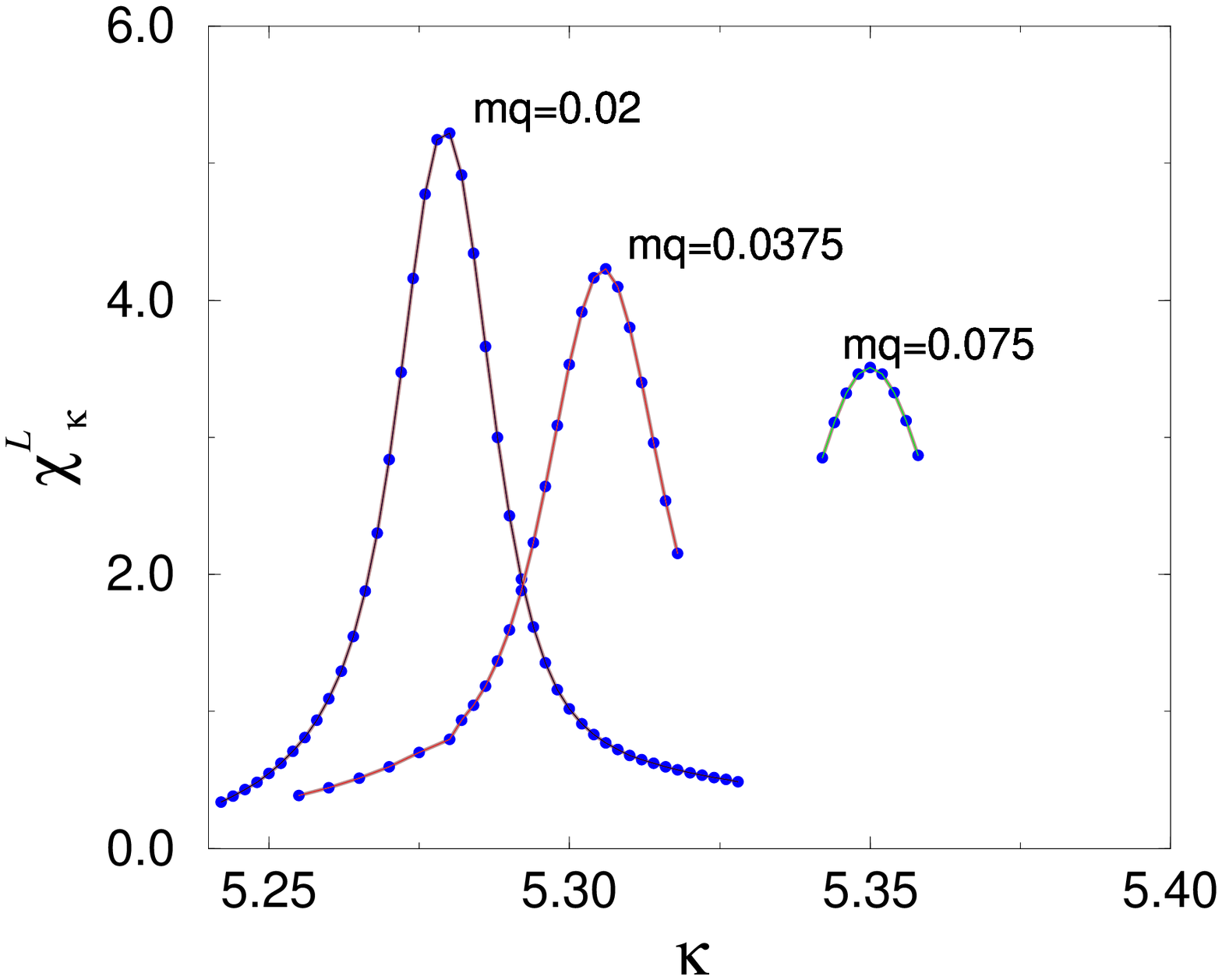,width=7cm}
\hskip0.5cm
\epsfig{file=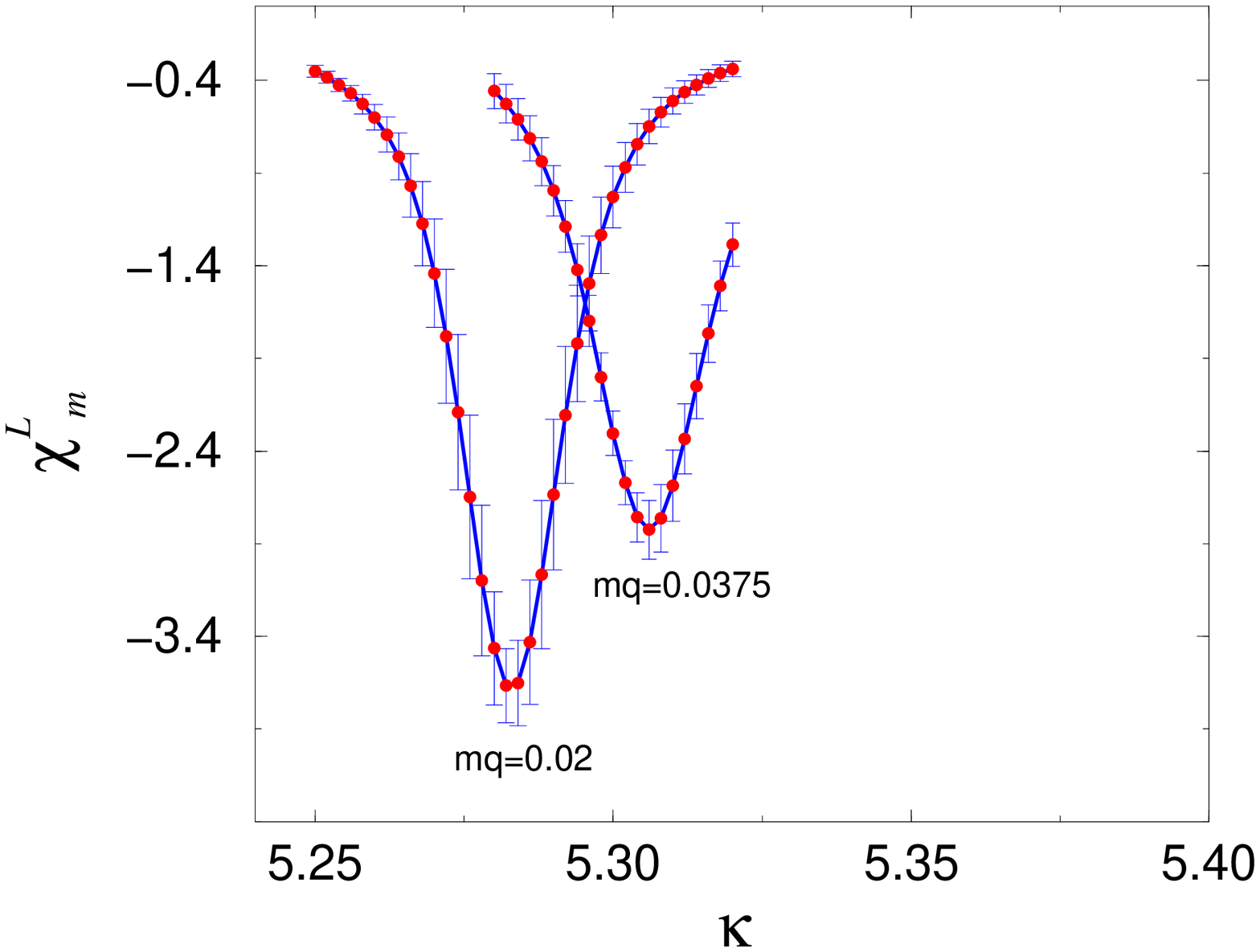,width=7cm}
\end{center}

\caption{The Polyakov loop susceptibilities $\x^L_{\kappa}$ and 
$\x^L_{\kappa}$ as functions of the temperature variable $\kappa=6/g^2$.}
\label{F2.3b}
\end{figure}

These considerations have immediate consequences which can be tested in
finite temperature lattice QCD. The magnetization $m(T,B)$ for
non-vanishing external field $B$ becomes an analytic function of $T$ and
$B$. In full QCD, we therefore assume the Polyakov loop for $m_q\not=0$
to be an analytic function of $T$ and $K$. This leads to
\be
dL = \left( {\partial L \over \partial T} \right)_K dT
+ \left( {\partial L \over \partial K} \right)_T dK.
\label{2.14a}
\ee
From this we obtain
\be
\x^L_m \equiv \left( {\partial L \over \partial m_q}\right)_T =
\left({\partial L \over \partial K}\right)_T \left({\partial K \over
\partial m_q}\right)_T,
\label{2.14}
\ee
and
\be
\x^L_T \equiv \left({\partial L \over \partial T}\right)_{m_q} =
\left({\partial L \over \partial K}\right)_T \left({\partial K
\over \partial T}\right)_{m_q}  +  \left( {\partial L \over \partial T}
\right)_K
\label{2.15}
\ee
for the Polyakov loop susceptibilities $\x^L_m$ and $\x^L_T$. Since the
chiral susceptibilities $\x^K_m =(\partial K/\partial m_q)_T$ and
$\x^K_T = (\partial K / \partial T)_{m_q}$ diverge at $T=T_{\x}$,
relations (\ref{2.14}/\ref{2.15}) imply that the Polyakov loop
susceptibilities must share this singular behaviour, with the same
critical exponents. Present lattice calculations for full QCD are not
yet precise enough to allow a conclusive determination of critical
exponents. In Figs.\ \ref{F2.3a} and \ref{F2.3b} it is seen, however, 
that the increase of the chiral susceptibilities $\x_m^K$ and $\x^K_T$
for $m_q\to 0$ is indeed accompanied by a similar increase in the Polyakov 
loop susceptibilities $\x^L_m$ and $\x^L_T$.
\medskip

In QCD with massless dynamical quarks, the chiral condensate $K$ and the
Polyakov loop $L$ thus are analytically related; there is only one
critical point $T=T_{\x}$, at which both $K(T)$ and $L(T)$ exhibit
non-analytic behaviour. At $T_{\x}$, the quarks loose their effective
mass and at the same time become unbound.

~~~\vskip1cm

\bigskip

\noindent{\bf \large 3.\ Colour Screening}

\bigskip

In pure $SU(N)$ gauge theory, the potential between a static quark and
antiquark increases linearly and unbounded with their separation $r$,
$V(r) \sim \sigma r$. In full QCD, the string breaks when $V(r) \simeq
m_Q$, even in vacuum, i.e., at $T=0$. We can interpret this by
attributing a screening behaviour to the sea of virtual massless
quark-antiquark pairs. Starting from Eq.\ (\ref{1.1}), we thus have
\be
V(r,\mu) = \sigma r \left[ {1 - e^{-\mu r} \over \mu r} \right]
- {\alpha \over r} e^{-\mu r},
\label{3.1}
\ee
where the last term includes Coulomb and transverse string effects. In
Eq.\ (\ref{3.1}), the string tension $\sigma$ and the Coulomb coupling
$\alpha$ are taken to be constants, with $\mu(T)$ temperature-dependent.
First, we now want to determine the vacuum screening mass $\mu(T=0)$.

\medskip

In the spectroscopy of heavy quarkonia, such as the \J~or the \U, the
masses and widths of all bound states are determined by solving the
Schr\"odinger equation
\be
{\cal H}\phi_i = M_i \phi_i,
\label{3.2}
\ee
with the Hamiltonian
\be
{\cal H} \equiv 2m_{c,b} - {\nabla^2 \over m_{c,b}} + V(r,\mu=0)
\label{3.3}
\ee
given in terms of the potential (\ref{3.1}). Here $i$ specifies the
$\C$ or $\B$ bound state under consideration, $M_i$ its mass.
By comparison for the results to quarkonium data, the four constants in
the potential are determined, giving $\sigma=0.192$ GeV$^2$,
$\alpha=0.471$ and $m_c=1.32$ GeV, $m_b=4.75$ GeV for the bare charm
and bottom quark masses, respectively. The vacuum screening mass can now
be obtained by comparing the gap between the open charm or bottom
threshold, $2M_D$ or $2M_B$, and a given quarkonium state $M_i$, to its
gap with respect to the infinite range potential,
\be
E_{\rm diss}^i = 2 M_{D,B} - M_i = 2m_{c,b} + {\sigma \over \mu} - M_i.
\label{3.4}
\ee
The result,
\be
\mu = {\sigma \over 2(M_{D,B} - m_{c,b}) }
\label{3.5}
\ee
gives $\mu(T=0) \simeq 0.18$ GeV for both charmonium and bottonium
states. The fact that the large difference between $m_c$ and $m_b$ plays
no role here is an indication that the states are indeed heavy enough to
estimate the medium effect alone. It is moreover reassuring that the
`Debye' screening length $r_D = \mu(T=0) \simeq 1.1$ fm is also the
expected hadronic scale.

\medskip

\begin{figure}[htb]
\centerline{\psfig{file=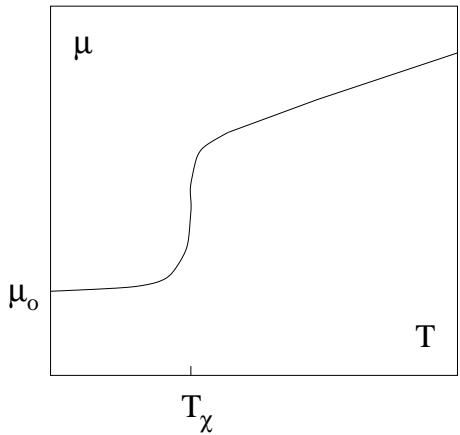,width=6cm}}
\vskip0.3cm
\caption{Schematic temperature dependence of the screening mass $\mu$
in full QCD with massless quarks of two flavours.}
\label{F3.2}
\end{figure}

At non-vanishing temperature, the screening mass can be determined
through a study of Polyakov loop correlations $\langle L(0) L^+(r)
\rangle$. Normalization problems make this non-trivial \cite{DLS}, so
that for the moment $\mu(T)$ is known only up to an open constant.
It is already clear, however, that $\mu(T)$ increases sharply around
$T=T_{\x}$, with $(\partial \mu / \partial T)$ diverging in the chiral
limit. Combining this result with the known
$\mu(T=0)$ and the perturbative result $\mu \sim gT$ leads to the
schematic screening mass form shown in Fig.\ \ref{F3.2}. In QCD, the
screening mass thus shows a very characteristic behaviour, with a
singular derivative at $T_{\x}$.

The physics leading to this phenomenon seems quite clear. Chiral
symmetry restoration transforms effectively massive into massless
quarks. Near $T=T_{\x}$, this sudden loss of mass leads to a sudden
increase in the density $n$ of coloured constituents and thus, with
$\mu \sim n^{1/3}$, also in the effectiveness of colour screening.

\medskip

An interesting side-line here is the onset of charmonium suppression as
signal for colour deconfinement \cite{Matsui}. Deconfined media will
dissociate charmonium and bottonium states in a step-wise pattern, with
the larger and more loosely bound states melting before smaller and more
tightly bound quarkonia. There are indications that the dissociation
point for the charmonium states $\x_c(1P)$ and \P$(2S)$ coincides with
$T_{\x}$ \cite{KMS}. If this could be substantiated in more precise
lattice studies, it would identify the onset of $\x_c$ and
\P~suppression with the onset of colour deconfinement.

\bigskip

\noindent{\large 4.\ Cluster Percolation}

\bigskip

Conceptually, the deconfinement transition seems rather transparent, no
matter what the quark mass is. Once the density of constituents becomes
so high that several hadrons have considerable overlap, there is no
mechanism to partition the quark constituents into colour-neutral bound
states. Instead, there appear clusters much larger than hadrons, within
which colour is not confined. This suggests that deconfinement is
related to cluster formation, and since that is the central topic of
percolation theory, possible connections between percolation and
deconfinement were discussed already quite some time ago on a rather
qualitative level \cite{Baym,Celik}. In the meantime, however, the
interrelation of geometric cluster percolation and critical behaviour
of thermal systems has become much better understood \cite{S-A}, and
this understanding can be used to clarify the nature of deconfinement.

\medskip

To recall the fundamentals of percolation, consider a two-dimensional
square lattice of linear size $L$; we randomly place identical objects
on $N$ of the $L^2$ lattice sites. With increasing $N$, adjacent
occupied sites will begin to form growing clusters or islands in the sea
of empty sites. Define $n_p$ to be the lowest value of the density $n =
N/L^2$ for which on the average the origin belongs to a cluster
reaching the edge of the lattice. In the limit $L \to \infty$, we then
have
\be
P(n) \sim \left(1 - {n_p \over n}\right)^{\beta_p},~~~n \geq n_p,
\label{4.1}
\ee
where the {\sl percolation strength} $P(n)$ denotes the probability
that the origin belongs to an infinite cluster. Since $P(n)=0$ for all
$n \leq n_p$ and non-zero for all $n > n_p$, it constitutes an order
parameter for percolation: $\beta_p=5/36$ is the critical exponent
which governs the vanishing of $P(n)$ at $n=n_p$ in two dimensions; in
three dimensions, it becomes $\beta=0.41$ \cite{Isi}.

\medskip

Another quantity of particular interest is the {\sl average cluster
size} $S(n)$, defined as the average number of connected occupied sites
containing the origin of the lattice; above $n_p$, percolating
clusters are excluded in the averaging. This cluster size corresponds to
the susceptibility in thermal systems and diverges at the percolation
point as
\be
\S(n) \sim |n - n_p|^{-\gamma_p}, \label{4.2}
\ee
with $\gamma_p=43/18$ (1.80) as the $d=2$ (3) critical exponent for the
divergence \cite{Isi}.

\medskip

We now turn once more to the Ising model.
For $B=0$, the Hamiltonian ({\ref{2.6})) has a global $Z_2$
invariance ($s_i \to -s_i~ \forall~ i$), and the magnetization $m =
\langle s \rangle$ probes whether this invariance is spontaneously
broken, as discussed in section 2. Such spontaneous symmetry breaking
occurs below the Curie point $T_c$, with
\be
m(T,B=0) \sim ( T_c - T )^{\beta_m}
\label{4.4}
\ee
governing the vanishing of $m(T,B=0)$ as $T \to T_c$ from below.
The well-known Onsager calculations with $\beta_m=0.125$ for $d=2$,
a value 10 \% smaller than the $\beta_p=5/36 \simeq 0.139$ found for the
percolation exponent.

\medskip

Since the Ising model also produces clusters on the lattice, consisting
of connected regions of aligned up or down spins, the relation between
its thermal critical behaviour at $T_c$ and the onset of geometric
percolation is an obvious question which has been studied extensively
in recent years. In other words, can one interpret magnetization
as spin domain fusion? This question is now answered \cite{F-K,C-K}.

\medskip

The geometric clusters in a percolation study consist of connected
regions of spins pointing in the same direction. In the Ising model,
there is a thermal correlation between spins on different sites; this
vanishes for $T \to \infty$. Correlated regions in the Ising model (we
follow the usual notation and call them `droplets', to distinguish them
from geometric clusters) thus disappear in the high temperature limit.
In contrast, the geometric clusters never vanish, since the probability
for a finite number of adjacent aligned spins always remains finite; it
increases with dimension because the number of neighbours does. Hence
from the point of view of percolation, there are more and bigger
clusters than there are Ising droplets.

\medskip

If percolation is to provide the given thermal critical behaviour, the
definition of cluster has to be changed such that the modified
percolation clusters coincide with the correlated Ising droplets
\cite{F-K,C-K}. This is achieved by assigning to pairs of adjacent
aligned spins in a geometric cluster an additional bond correlation,
present with the bonding probability
\be
p_b = 1 - \exp\{-2J/kT\}, \label{4.5}
\ee
where $2J$ corresponds to the energy required for flipping an aligned
into a non-aligned spin. The modified `F-K' percolation clusters now
consist of aligned spins which are also bond-connected. Only for $T=0$
are all aligned spins bonded; for $T >0$, some aligned spins in a purely
geometric cluster are not bonded and hence do not belong to the
modified cluster or droplet. This effectively reduces the size of a
given geometric cluster or even cuts it into several modified clusters.
For $T \to \infty$, $n_b \to 0$, so that the geometric clusters still
in existence there are not counted as droplets, solving the problem
mentioned in the previous paragraph.

\medskip

For such combined F-K site/bond clusters, full agreement between
percolation and thermal critical behaviour of the Ising model
is achieved for any space dimension $d$. The percolation threshold is
now at $T_c$, the cluster size coincides with that of the correlated
regions in the Ising model, and numerical simulations show that the
critical exponents for the new cluster percolation scheme become those
of the Ising model.

\medskip

Since, as noted, the deconfinement transition in $SU(N)$ gauge theory
falls into the universality class of the Ising model, it seems natural
to look for a percolation formulation of deconfinement \cite{HS-deco},
and first studies indicate that this is indeed possible
\cite{Santo1,Santo2}. In $SU(2)$ lattice gauge theory, the Polyakov loop
constitutes essentially a generalized spin variable, pointing either up
or down at each spatial lattice site, but with continuously varying
magnitude. In two space dimensions, this leads to a `landscape' of hills
and lakes of various heights and depths. The crucial question in the
extension of percolation to such a case is how to generalize the bond
weight Eq.\ (\ref{4.5}). For a specific lattice regularization, the
strong coupling limit, it was shown that the action in $SU(2)$ gauge
theory can effectively be written in terms of nearest neighbour Polyakov
loop interactions, with $(\kappa/4)^2 L_iL_j$ in place of the Ising
form $(J/kT)s_is_j$, where $\kappa\equiv 4/g^2$ and $g$ denotes the
coupling in the gauge theory action \cite{G-K}. We therefore take
\be
p^b_{i,j} = 1 - \exp\{-2(\kappa/4)^2 L_iL_j\}
\label{4.6}
\ee
as bond weight between two adjacent Polyakov loops of the same sign.
It is known through analytic as well as numerical studies that such a
form gives the correct bond weight for continuous spin Ising models
\cite{Gandolfo,Santo3}. To test it here, we have carried out studies
on a number of different lattices for both two and three space dimensions;
some results are shown in Fig.\ \ref{F4.1} and in Table 1. Fig.\
\ref{F4.1} shows that the rescaled percolation probability, using the
Ising value for the exponent
$\nu$ leads to a universal curve, as required. In the table we
summarize the excellent agreement between thermal and percolation values
for the critical exponents. The exponents for random site percolation,
on the other hand, are seen to be considerably different.

\begin{table}[htb]
\begin{center}
\label{Table 1}
\newcommand{\m}{\hphantom{$-$}}
\newcommand{\cc}[1]{\multicolumn{1}{c}{#1}}
\renewcommand{\tabcolsep}{1pc} 
\renewcommand{\arraystretch}{1.2} 
\begin{tabular}{@{}llll}
\hline
&\cc{${\beta}/{\nu}$} &
\cc{${\gamma}/{\nu}$} & \cc{$\nu$} \\
\hline
L-Percolation
&0.528(15)
&1.985(13)  & 0.632(11) \\
Symmetry Breaking
&0.523(12)
& 1.953(18)& 0.630(14) \\
Ising Model \cite{ferr} & 0.518(7)  & 1.970(11) & 0.6289(8)  \\
Random Percolation \cite{Balle} &  0.4770(10)  &  2.0460(39)  &  
0.8765(16)  \\
\hline
\end{tabular}\\[2pt]
\end{center}
\caption{Comparison of percolation and thermal exponents for
3+1 $SU(2)$.}
\end{table}

\bigskip

\begin{figure}[htb]
\vskip0.3cm
\centerline{\psfig{file=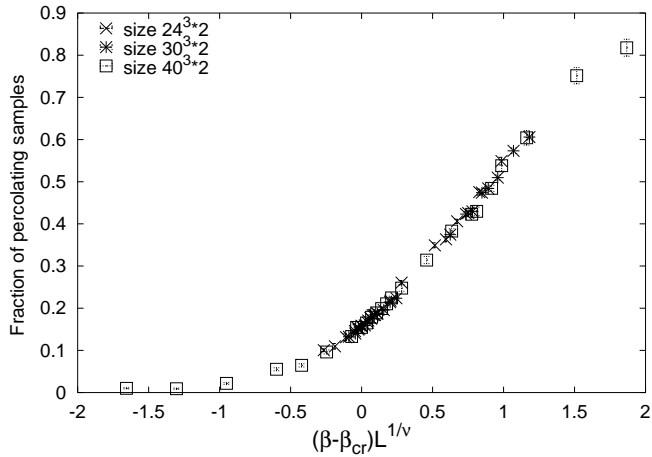,width=9cm}}
\caption{Rescaled percolation probability for 3+1 $SU(3)$ gauge theory,
using the Ising exponent $\nu=1$.}
\label{F4.1}
\end{figure}

For the specific lattice regularization used, we thus indeed find that
deconfinement in $SU(2)$ gauge theory can be described as Polyakov loop
percolation. Hopefully this can be extended to more general lattice
regularizations.

\medskip

The greatest interest in a percolation approach to deconfinement is,
however, based on the possibility to define the transition for arbitrary
values of the quark mass. Consider the case of full QCD with colour
$SU(3)$ and two massless quark flavours. For $m_q \to \infty$, this
leads to the first order transition of $SU(3)$ gauge theory, as
counterpart of such a transition in the three-state Potts model. For
decreasing quark mass, the transition will eventually disappear at a
second-order end point defined in terms of $m_q^c$, and for $m_q^c > m_q
>0$, there presumably is no thermal transition at all. Finally, for
$m_q \to 0$, there is the chiral symmetry restoration transition. In
Fig.\ \ref{F4.2}, this behaviour is illustrated. Does this mean that in
the quark mass region $m_q^c > m_q >0$ (which includes our real
physical world of small but finite bare quark masses) there is
no way to define deconfinement as a critical phenomenon?

\begin{figure}[htb]
\centerline{\psfig{file=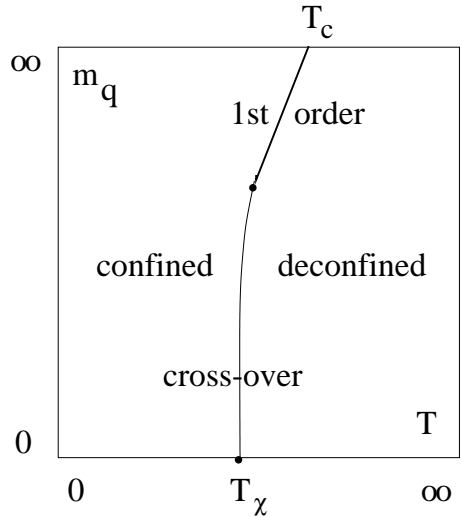,width=6cm}}
\caption{QCD phase structure as function of temperature $T$
and quark mass $m_q$.}
\label{F4.2}
\end{figure}

\medskip

To address this question, we return to percolation in the Ising model
with a non-vanishing external field. For $H \not= 0$, the Ising
partition function does not contain any singularity as function of $T$
and hence does not show any critical behaviour \cite{L-Y}; the $Z_2$
symmetry responsible for the onset of spontaneous magnetization is now
always broken and $m(T,H\not=0) \not= 0$ for all $T$. On the other
hand, the average size of site/bond clusters in the above sense
increases with decreasing temperature, and above some critical
temperature it diverges. Hence percolation will occur for any
value of $B$. In other words, the critical behaviour due to percolation
persists, while that related to spontaneous symmetry breaking and
magnetization disappears. At $B=\infty$, all spins are aligned, leaving
the bonds as the relevant variables; the system now percolates at the
critical density for pure bond percolation, which leads to a
critical temperature $T_k$ somewhat above the Curie point $T_c$.
For finite $B$, the corresponding values of the critical temperature
lie between $T_c$ at $B=0$ and $T_k$ at $B=\infty$; they define the
so-called Kert\'esz line \cite{S-A,Kertesz}; see Fig.\ \ref{F4.3}.
A fundamental and quite general question in statistical physics is
what happens at this line. Can one generalize critical behaviour
to situations where the partition function $Z(T)$ is analytic, but where
percolation as defined in terms of the input dynamics persists? It is
evident from the similarity of Figs.\ \ref{F4.2} and \ref{F4.3} that the
answer is immediately relevant to the study of phase transitions in QCD,
with deconfinement as the QCD counterpart of the Kert\'esz line.

\begin{figure}[htb]
\centerline{\psfig{file=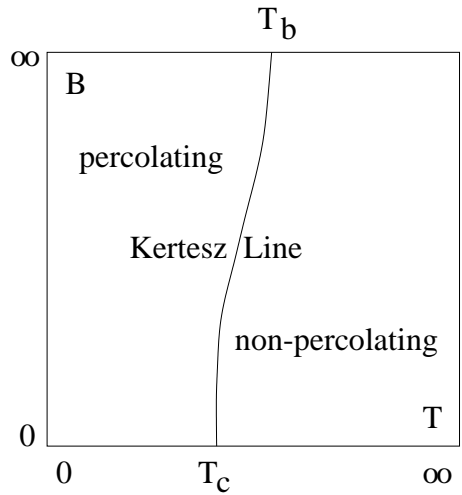,width=6cm}}
\caption{Percolation pattern for the Ising model with external
magnetic field $B$.}
\label{F4.3}
\end{figure}

\medskip

In statistical physics, such generalized critical behaviour indeed leads
to observable effects. Familiar instances are found in solution/gel
transitions, as encountered in the boiling of an egg or the making of
pudding. While these phenomena do not result in any thermodynamic
singularities, they are well-defined transitions which can be
quantitatively described in terms of percolation \cite{Coniglio}.

\medskip

In closing, we note another possible illustration of the relation
between percolation and thermal critical behaviour. Consider the F-K
site/bond percolation in the two-dimensional Ising model as introduced
above, and imagine that current can flow between two or more bonded
sites. In this case, conductivity sets in at the percolation point,
independent of the magnetization critical behaviour of the Ising model;
the system is non-conducting below the percolation point and conducting
above it. In other words, we now have two independent critical
phenomena, the onset of conductivity and the critical behaviour of Ising
thermodynamics, and the former can survive even when the latter is no
longer present.

\bigskip

\noindent{\large 5.\ Summary}

\bigskip

We have seen that hadronic matter at sufficiently high temperature
and low baryon density becomes a quark-gluon plasma. In this
deconfinement transition, the colour-neutral bound-state constituents
are dissolved into their coloured components. At high baryon density
and low temperature, the deconfined medium could be a condensate of
coloured bosonic diquarks.

\medskip

In pure $SU(N)$ gauge theory, colour deconfinement arises through the
spontaneous breaking of a global $Z_N$ symmetry of the Lagrangian. In
the chiral limit of full QCD, it occurs through a strong explicit
breaking of this symmetry, due to an effective external field setting in
when the chiral condensate vanishes. Hence the deconfinement and chiral
symmetry restoration transitions coincide.

\medskip

Colour charge screening in QCD leads to a specific singular behaviour of
the screening mass. At the critical temperature of chiral symmetry
restoration, the effective quark mass shift leads to sudden increase in
the density of constituents and hence to more effective screening. A
particularly enticing question here is whether the dissociation of the
$\x_c$ state occurs at just this point - it would then be a measurable
order parameter for deconfinement.

\medskip

Cluster percolation provides an approach to study the geometry of
deconfinement. In $SU(N)$ gauge theory, first finite temperature
lattice calculations indicate that Polyakov loop percolation is indeed
an equivalent way to identify deconfinement. In full QCD as in spin
systems with non-vanishing external field, percolation persists even
in the absence of thermal transitions. It thus seems conceivable to
identify the colour deconfinement transition for arbitrary quark mass
through the onset of percolation.

\bigskip

\noindent{\large Acknowledgements}

\bigskip

It is a pleasure to thank Ph.\ Blanchard, S.\ Digal, S.\ Fortunato,
D.\ Gandolfo, F.\ Karsch, E.\ Laermann and P.\ Petreczky for helpful
discussions on different aspects of this survey.

\end{document}